\documentclass[12pt]{iopart}
\usepackage{graphicx}
\begin{document}

\title[Scaling and anomalous fluctuation statistics]{Scaling and commonality in anomalous fluctuation
 statistics in  models for turbulence and ferromagnetism.}
\author{S C Chapman\dag \ddag, G Rowlands\dag and N W Watkins$\sharp *$}
\address{\dag Space and Astrophysics, University of Warwick, UK}
\address{\ddag also at the Radcliffe Institute for Advanced Study,
Harvard University, UK}

\address{$\sharp$ British Antarctic Survey (NERC), Cambridge, UK}
\address{* on leave at Center for Space Research, Massachusetts Institute of Technology,
Cambridge, Mass., USA}

\date{\today}
\maketitle
\begin{abstract}
Recently, Portelli et al (2003) have semi- numerically obtained a
functional form of the probability distribution of fluctuations in
the total energy flow in a model for fluid turbulence. This
follows earlier work suggesting that fluctuations in the total
magnetization in the 2D X-Y model for a ferromagnet also follow
this distribution. Here, starting from the scaling anzatz that is
the basis of the turbulence model we analytically derive the
functional form of this distribution and find its single control
parameter that depends upon the scaling exponents and system size
of the model. Our analysis allows us to identify this explicitly
with that of the X-Y model, and suggest a possible generalization.
\end{abstract}


\section{Introduction}

Scaling is an important feature of natural phenomena, arising in
many degree of freedom systems that are highly correlated. In
reality these systems support a finite range of scales, from the
microscopic to the system size. If the number of degrees of
freedom is sufficiently large, these systems will still fall into
the framework of critical phenomena \cite{goldenfeld}. Such
``inertial" \cite{bramprl} systems include a disparate range of
phenomena and, non-intuitively, have recently been suggested
 to have a common signature in the
statistics of fluctuations in  global measures of activity. This
probability distribution function (PDF) has been compared
numerically \cite{bramprl} for a range of models including some
for out of equilibrium critical phenomena, notably a sandpile, a
forest fire model, a depinning model and a stacking model for
granular media. After normalization to the first two moments these
PDF were found to collapse onto that of fluctuations in models for
equilibrium critical phenomena.

The functional form of this curve has been identified
semi-numerically \cite{brampre} with the distribution
\cite{bramprl}:
\begin{eqnarray}\label{gumb}
P(y)= K e^{a_g(u-e^u)},\;\;\;\; u=b(y-s)
\end{eqnarray}
where the constants $K,b$ and $s$ are fixed by setting the moments
$M_0=1, M_1=0$ and $M_2=1$, leaving a single  parameter, $a_g$.

The statistics  of fluctuations in global quantities have been
explored both experimentally and theoretically for fluid
turbulence in closed systems. In the experiments of Labb\'{e}
\emph{et al}. \cite{pintonlabbe,pintonh},
 the normalized PDF of
fluctuations in the power provided to both rotor blades stirring a
closed cylinder of gas at constant angular frequency was also
found to have collapse onto (\ref{gumb}) over a range of Reynolds
number $R_e$. These results have been compared with the PDF of
fluctuations in the total magnetization in the 2D X-Y model for a
ferromagnet \cite{bramnature} which also has been identified with
(\ref{gumb}) \cite{brampre}. In the experiment reported by
\cite{cadot} (their Figure 2), the normalized PDF of fluctuations
in wall pressure, rather than injected power, appear to follow
these non- Gaussian statistics with insensitivity to Reynolds
number \cite{cadot}. Recently, \cite{portellipp} treated a model
for closed turbulence and obtained a family of curves of the form
(\ref{gumb}) semi- numerically for global fluctuations of kinetic
energy which within a sign the experimentally measured total power
is believed to act as a good proxy \cite{portellipp}.

There is a continuing debate (see
\cite{aji,zhengcomment,bramzreply,dahlstedt,ncomment,bramnreply,gyorgi,clousel} and references therein)
concerning the origin of this apparent universality
\cite{bramnature,bramprl}. The aim of this paper is not to
establish the existence, or lack thereof, of a universality class.
Furthermore, we do not address the correspondence between
experimentally measured quantities and those captured by models
for turbulence (see \cite{portellipp}).

Here, we analytically derive (\ref{gumb}) for the model for
intermittent turbulence in a finite sized system treated by
\cite{portellipp}. We obtain $a_g$ as a function of the model
parameters. The analysis then leads to a direct identification
with results obtained previously for the 2D X-Y model
\cite{archemb}, elucidating the origin of the value $a_g\sim
\pi/2$ obtained for that system \cite{bramnature,bramprl}.  We
then suggest that the features of the model that are intrinsic to
this calculation are
 rather generic and discuss how they may
encompass the wide variety
 of systems which have also been previously identified as
 exhibiting  the same functional form  for the fluctuation PDF \cite{bramprl}.

\section{Model for turbulence in a finite sized system}
 Portelli \emph{et al.}
\cite{portellipp} obtained (\ref{gumb}) semi-numerically for
intermittent turbulence in the framework of the KO62 hypothesis
which  models fluctuations in the energy in the flow.

 The essential features of this model are structures on a range of  length scales
$l_1..l_j..l_N$ from a smallest size $l_1=\eta$ to the system size
$l_N=L$, corresponding to the dissipation and driving length
scales respectively. The Reynolds
  number of the flow is then
defined as $R_e=(L/\eta)^{4/3}$ \cite{frisch}, and the ratio
between successive lengthscales
$(l_j/l_{j-1})=\lambda^\frac{1}{3}$ so that
$\lambda^N=(L/\eta)^3$.

Following \cite{portellipp} we wish to calculate the statistics of
the total energy in the flow:
\begin{equation}
\varepsilon(t)=\epsilon_1(t)+\epsilon_2(t)+\cdots+\epsilon_j(t)\cdots+\epsilon_N(t)\label{esum0}
\end{equation}
from a model expressed in terms of an intermittent energy transfer
rate $\epsilon_j$ on lengthscale $l_j$ drawn from a PDF which has
moments:
\begin{equation}
<\epsilon_j^q>
 =\epsilon_0^q\left(\frac{l_j}{L}\right)^{-\mu(q)}\label{scale}
\end{equation}
with the condition $<\epsilon_j>=\epsilon_0$ which fixes
$\mu(1)=0$. It follows that the standard deviation:
\begin{equation}
\sigma_j^2=\frac{\epsilon_0^2}{\nu_j}
=\epsilon_0^2\left[(l_j/L)^{-\mu(2)}-1\right]\label{sigj}
\end{equation}

The individual $\epsilon_j$ are assumed independent, each with PDF
$P_j(\epsilon_j)$, giving:
\begin{eqnarray}\label{ptotal}
P(\varepsilon)=\int\delta(\varepsilon-\sum^N_{j=1}\epsilon_j)\prod^N_{l=1}P_l(\epsilon_l)d\epsilon_l\\
=\frac{1}{2\pi}\int^\infty_{-\infty}e^{ik\varepsilon}dk\prod_{l=1}^N\hat{P_l}(k)\nonumber
\end{eqnarray}
where $\hat{P}_l(k)=\int P_l(\epsilon_l)d\epsilon_l
\exp(-ik\epsilon_l)$.

In the framework of KO62  this scaling system supports a cascade
from large to small scales, with  the intermittency parameter
$\tau(2)=-\mu(2)$. Importantly, although we can envisage a
cascade, the above model does
 not explicitly
require one and will map onto other models provided that the basic
assumptions, namely of $\epsilon_j$ that are independent and drawn
from a PDF with the scaling property (\ref{scale},\ref{sigj}),
hold.

The $P_j$ will depend upon the details of the system, to make
progress we first consider a tractable choice, the $\chi^2$
distribution:
\begin{equation}
P_j(\epsilon_j)=A_j
\epsilon_j^{\nu_j-1}e^{-a_j\epsilon_j}\label{kisq}
\end{equation}
and later explore how this may be generalized. The influence of
the microscopic distribution $P_j$ has been explored in the
context of the 2D X- Y model in \cite{phxy}. This choice for
$P(\epsilon_j)$ was used to evaluate (\ref{ptotal}) semi
numerically in \cite{portellipp}; we will now evaluate it
analytically and as a corollary directly obtain the solution
previously obtained semi- numerically for the 2D XY model
\cite{brampre}.

\section{Evaluating the Characteristic Function}

The constants $A_j, \nu_j$ and $a_j$ of (\ref{kisq}) are fixed
through  (\ref{sigj}) so that Eq. (\ref{ptotal}) can be written as
\cite{portellipp}:
\begin{eqnarray}
P({\varepsilon})=\frac{1}{2\pi}\int^\infty_{-\infty}
e^{ik\varepsilon}dk e^{-S_N} \;\;\;\;\;\textrm{where}\\
\label{pint} S_N=\sum_{j=1}^N\frac{1}{f_j}\ln( 1+ik\epsilon_0 f_j
)=\sum_{j=1}^N \frac{1}{f_j}\ln(1+ikf_j \frac{\beta \sigma
}{N})\label{sn}
\end{eqnarray}
where \begin{equation} f_j=1/\nu_j=\exp(j\bar{a})-1
\end{equation}
$\bar{a}=(\mu(2)/3)\ln\lambda$ and where we define total variance
\begin{equation}
\sigma^2=\sum^N_{j=1}\sigma_j^2=\epsilon_0^2\sum_{j=1}^N f_j
\end{equation}
and $\beta=N\epsilon_0/\sigma$. For a specific system, if one
has the values of $\mu(2)$ and $\lambda$, and that (\ref{kisq}) is
a good approximation for $P_j(\epsilon_j)$, one can evaluate
(\ref{pint}) numerically \cite{portellipp}. Here, however, we wish
to establish why (\ref{gumb}) appears to also describe the 2D X-Y
model. We proceed by analytically evaluating (\ref{pint}), and
begin with $S_N$.
 By formal expansion:
\begin{eqnarray}\label{dsndk}
\frac{dS_N}{dk} =i\frac{\beta \sigma }{N}
\sum_{n=0}^\infty\left(-ik\frac{\beta \sigma
}{N}\right)^n\sum_{j=1}^Nf_j^n
\end{eqnarray}
We need to find $F_n=\sum_{j=1}^Nf_j^n$ with the condition that
$F_0=N$ and $F_1=\sum_{j=1}^Nf_j=(N/\beta)^2$. From our definition
of the $f_j$ we can evaluate $F_1$ and obtain
\begin{equation}
\exp(N\bar{a})=1+(N+N^2/\beta^2)(1-\exp(-\bar{a})) \end{equation}
In the limit of $N/\beta^2>1$, that is, $N\gg 1$, so that
$\mid\bar{a}\mid\ll 1$ for finite system size $L$ (ie $\lambda
\rightarrow 1$) this gives:
\begin{equation}
e^{N\bar{a}}\approx 1+\frac{N^2 \bar{a}}{\beta^2}\label{nacond}
\end{equation}
We can now expand the $F_n$ in $e^{N\bar{a}}$ and substitute
(\ref{nacond}):
\begin{eqnarray}
F_n=\sum_{j=1}^N (e^{j\bar{a}}-1)^n=\sum_{j=1}^N
e^{nj\bar{a}}-n\sum_{j=1}^Ne^{(n-1)j\bar{a}}+\ldots\\\nonumber
=\frac{e^{nN\bar{a}}-1}{1-e^{-n\bar{a}}}-n\frac{e^{(n-1)N\bar{a}}}{1-e^{-(n-1)\bar{a}}}+\ldots\\\nonumber
\approx
\frac{N^n}{n\bar{a}}\left(\frac{N\bar{a}}{\beta^2}\right)^n
 +{\cal
 O}\frac{N^{n-1}}{n\bar{a}}\left(\frac{N\bar{a}}{\beta^2}\right)^{n-1}+\ldots
 \end{eqnarray}
 and in our limit $N\gg 1$, $\mid\bar{a}\mid\ll 1$ such that $N/\beta^2>1$
 we can take $N\mid\bar{a}\mid/\beta^2\sim 1$ to give to lowest order
 $F_n\approx (N^n/(n\bar{a}))((N\bar{a})/\beta^2)^n$
 where $n \geq 1$ ($F_0=N$). Using this in
 (\ref{dsndk}) gives
 \begin{equation}
 \frac{dS_N}{dk}
 =i\sigma\beta-
 i\sigma {\cal Z} \ln(1+ik\sigma/\cal Z)
 \end{equation}
 where ${\cal Z} =\beta/(N\bar{a})$.
 This now readily integrates to give:
 \begin{eqnarray}
 S_N(k)=ik\sigma\beta +\psi(k)\;\;\;\;\textrm{where}\\
\psi(k)=ik\sigma {\cal Z}-{\cal Z}^2\left(1+ik\frac{\sigma}{\cal
Z}\right) \ln \left(1+ik\frac{\sigma}{{\cal Z}}\right)\label{epsi}
 \end{eqnarray}
 With $\phi=\varepsilon-\epsilon_0 N=\varepsilon-\beta \sigma$
 we can write (\ref{pint}) as:
 \begin{eqnarray}
 P(\varepsilon)=
 \frac{1}{2\pi}\int^{\infty}_{-\infty}
 e^{ik\phi}e^{-\psi (k)} dk\label{pehalf}
 \end{eqnarray}
 The limit $N\bar{a}\rightarrow 0$, ${\cal Z}\rightarrow \infty$,
corresponds to retaining terms up to $k^2$ in (\ref{epsi}) and
immediately gives a Gaussian distribution for $P(\varepsilon)$. It
is tempting to take this as the Gaussian limit of
$P(\varepsilon)$, $\bar{a}\rightarrow 0$, or $\mu(2)\rightarrow
0$. However, since we insisted that the PDF $P_j(\epsilon_j)$  are
scaling, this limit would yield $\sigma_j\rightarrow 0$ in
(\ref{sigj}). The case where the $P_j(\epsilon_j)$ all have the
same (that is, non-scaling) $\sigma_j$, corresponding to $N
\bar{a} \rightarrow 0$ as this gives $F_{2..N}\rightarrow 0$
above. To evaluate (\ref{pehalf}) in general we need to retain the
property of scaling $\sigma_j$, thus excluding this limit.

We evaluate (\ref{pehalf}) by the method of steepest descent:
\begin{equation}
P(\varepsilon)=\frac{e^{{\cal Z}^2}}{\sigma \sqrt{2\pi}}
e^{-{\frac{\phi}{\sigma}}({\cal Z}+\frac{1}{2{\cal Z}})-{\cal Z}^2
e^{-\frac{\phi}{\sigma Z}}}
\end{equation}
which is of the form (\ref{gumb}) with $a_g={\cal Z}^2+1/2$, that
is,
\begin{equation}
a_g=\frac{1}{2}+\frac{1}{\bar{a}\left(e^{N\bar{a}}-1\right)}\label{agumb}
\end{equation}
and $u=-\varepsilon (\bar{a}/\epsilon_0)+(N\bar{a}+A)$ with $A$
just given by the normalization constant $K$.

The present choice of $P_j(\epsilon_j)$, Eq. (\ref{kisq}), gives
exactly:
\begin{equation}
\hat{P_j}(k)=\frac{1}{(1+ik\kappa_j)^{\gamma_j}}\label{pade}
\end{equation}
with the $\kappa_j$ and $\gamma_j$ fixed by the moments. Other
forms of $\hat{P_j}(k)$ for which (\ref{pade}) is a good (Pad\'{e}
type) approximant will yield a $P(\varepsilon)$ of the form
(\ref{gumb}). We write (\ref{pade}) as:
\begin{eqnarray}
\hat{P_j}(k)=\sum_{p=0}^{\infty}\frac{(-ik)^p}{p
!}<\epsilon^p_j>\\\nonumber
 \simeq
1+ik\epsilon_0-\frac{k^2\epsilon_0^2}{2}(1+\frac{1}{\gamma_j})+\ldots
\nonumber
\end{eqnarray}
given that $<\epsilon_j>=\epsilon_0=\kappa_j\gamma_j$. All the
coefficients in this expansion are fixed if we insist that any
$P_j$ that we consider has the same scaling (\ref{sigj}) for
$\sigma_j$, so that $\exp(j\bar{a})=(1+1/\gamma_j)$. For the
$\chi^2$ PDF (\ref{kisq}) this gives $\gamma_j=\nu_j$ which is
exact. For example, one might anticipate that in a correlated
system that local fluctuations may be multiplicative. An
appropriate model for multiplicative noise is a lognormal PDF:
\begin{equation}
P_j(\epsilon_j)=\frac{1}{\sqrt{2\pi}}\frac{1}{\bar{\sigma}_j\epsilon_j}
\exp\left[-\frac{\left(\ln(\frac{\epsilon_j}{\bar{\epsilon}_j})\right)^2}{2\bar{\sigma}_j^2}\right]
\end{equation}
which has Pad\'{e} type approximant of form (\ref{pade}) fixed by:
\begin{equation}
<\epsilon_j^p>=\bar{\epsilon}^pe^{\frac{1}{2}p^2\bar{\sigma_j}^2},\;\;\;\;1+\frac{1}{\gamma_j}=e^{\bar{\sigma_j}^2}
\end{equation}
Thus if this is a good approximant, the lognormal also yields a
curve of form (\ref{gumb}).

\section{Results for the turbulence model.}

 To make a direct comparison with the results of \cite{bramnature,portellipp}
 we write
(\ref{agumb}) in terms of the parameters relevant to the
turbulence model:
\begin{equation}
a_g=\frac{1}{2}+ \frac{3}{\mu(2)
\ln(\lambda)}\left(R_e^{\frac{3\mu(2)}{4}}-1\right)^{-1}\label{ag}
\end{equation}
so that $a_g$ depends weakly on the Reynolds number of the flow,
on (experimentally determined) $\tau(2)$  and through the
logarithm, on the free parameter $\lambda$. In Figure 1 we show
normalized curves for the range $R_e=[10^4,10^5, 10^6]$
corresponding to $a_g=[7.7..3.6]$ explored by \cite{pintonh},
$\lambda=2$ \cite{portellipp} and typical values of $\tau(2)$
\cite{frisch}. Curves for $\tau(2)=-0.25$ are shown in the main
plot and  for $\tau(2)=-0.2$  (used \cite{portellipp}) in the
inset. In both cases these show good correspondence with the
solutions from the model (see Figures 1 and 2 of
\cite{portellipp}). The curves also fall close to each other
explaining the relative insensitivity to Reynolds number
\cite{portellipp}, and also fall close to that for $a_g=\pi/2$
identified by \cite{bramnature} for the X-Y model, which is also
plotted and which we discuss next.

\begin{figure}
\begin{center}
\noindent\includegraphics[width=20pc]{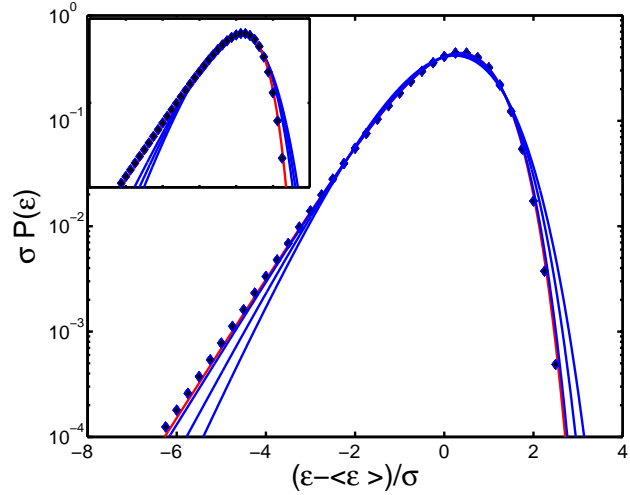}

 \caption{Function (\ref{gumb}) is plotted for $a_g=[4.24,2.76,1.90]$
(blue lines), $a_g=1.74$ (red line) and $a_g=\pi/2$ (diamonds)
relevant to the model \cite{portellipp} for a closed turbulence
experiment,
 the 2D X-Y model and \cite{bramprl} respectively.
The inset shows the same axis ranges and replaces the blue curves
with $a_g=[7.76, 5.18, 3.62]$ relevant to \cite{portellipp}. The
handedness is chosen for comparison with
\cite{bramprl,bramnature,portellipp}.}
 \end{center}
\end{figure}

A more sensitive method for identifying $a_g$ from data is to
calculate the third moment of (\ref{gumb}) directly \cite{chapnpg}
rather than compare curves. With $M_0=1, M_1=0, M_2=1$:
\begin{equation}
M_3=-\frac{N\bar{a}}{\beta}=-\sqrt{-\frac{\tau\ln\lambda}{3}\left(R_e^{-\frac{3\tau}{4}}-1\right)}
=-\left(a_g-\frac{1}{2}\right)^{-\frac{1}{2}}\label{m3}
\end{equation}
so its magnitude  increases with decreasing $a_g$ which from
(\ref{ag}) corresponds to increasing $R_e$, consistent with data.
The handedness is not determined here; this depends upon how the
global quantity measured experimentally relates to that of the
model; namely the energy flow within the fluid \cite{portellipp}.

 \section{The 2D X-Y Model.} The total
magnetization in the spin-wave approximation to the 2D XY model
was also shown for a range of system sizes \cite{bramnature} to
collapse onto (\ref{gumb}), with $a_g \simeq \pi/2$ (for a study
of the full 2D XY model see \cite{palma}). The PDF of total
magnetization $M$ can be written as \cite{archemb}:
\begin{eqnarray}\label{xy}
Q(M)&=&\int^{\infty}_{-\infty} \frac{dt}{2\pi\sigma}
e^{-it\frac{M-<M>}{\sigma}} e^{{\cal S}}\\
 \textrm{where}\;\;
{\cal
S}&=&\sum_{p=2}^{\infty}\left(-it\sqrt{\frac{2}{a_2}}\right)^p\frac{a_p}{2p},\;\;\;\;
\nonumber a_p=\frac{1}{N_s^p}\sum_{q\neq 0}\frac{1}{\gamma_q^p}
\end{eqnarray}
and $\gamma_q$ specifies the lattice Green's function. Bramwell et
al \cite{brampre} evaluated this numerically to demonstrate that
it is well described by (\ref{gumb}). We now show that the sum
${\cal S}$ is related to the sum $S_N$ (\ref{sn}) by writing:
\begin{eqnarray}
-S_N=-\sum^{N}_{p=1}\frac{\ln(1+ik\epsilon_0f_p)}{f_p}
=\sum_{p=1}^{N}\frac{1}{f_p}\sum^{\infty}_{m=1}\frac{(-ik\epsilon_0f_p)^m}{m}
\\\nonumber
=\sum^{\infty}_{m=1}\frac{(-ik\epsilon_0)^m}{m}F_{m-1}=-it<\epsilon>+\sum^{\infty}_{p=2}\frac{(-ik\epsilon_0)^p}{p}F_{p-1}
\nonumber
\end{eqnarray}
where $<\varepsilon>=N\epsilon_0$. If we then make the
identification:
\begin{equation}
F_{p-1}\epsilon_0^p\equiv\frac{a_p}{2}\left(\sqrt{\frac{2}{a_2}}\right)^p\label{ident}
\end{equation}
then $Q(M)$ (\ref{xy}) has the same functional form as
$P(\varepsilon)$ so that it also shares the same distribution
(\ref{gumb}) to within the approximations made here, namely, that
following expansion (14) we have neglected terms of order $1/N$,
and that (27) is also an approximation, good for $N$ large.
Importantly, from the definition of $a_p$ in (\ref{xy}), the
r.h.s. of (\ref{ident}) is independent of $N_s$. In this sense we
have approximately evaluated the integral (\ref{xy}) in the
thermodynamic limit.

It then just remains to estimate $a_g$ for the 2D X-Y model. In
\cite{brampre} this was approximated asymptotically, here we
simply note that insisting that the normalized $Q(M)$ and
$P(\varepsilon)$ share the first three moments yields
$a_g=1/2+(a_2/2)^3/a_3^2$ from (\ref{m3}) and equation (21) of
\cite{brampre}. They also calculated the normalized third moment
for a square lattice. Their value of $M_3=-0.8907$ gives
$a_g=1.7428$ which will give curves close to those for $a_g=\pi/2$
 as shown on Figure 1. Our analysis is thus consistent with both a
 value of $a_g \approx \pi/2$
 \cite{bramprl}, and the asymptotic exponent of \cite{brampre}.
\section{Generalization}
A variety of  disparate systems  have recently been shown
numerically \cite{bramnature,bramprl} to have a common signature
in the statistics of fluctuations in a global measure of activity
which is of the form (\ref{gumb}). These include  out of
equilibrium critical phenomena, notably a sandpile, a forest fire
model, a depinning model and a stacking model for granular media.

We will now argue that the scaling anzatz which was our starting
point for the model for fluid turbulence in section 2 and our
derivation of (\ref{gumb}) may also encompass these disparate
systems.

 The ansatz we
chose corresponds to that of a scaling system that generates
spatial structures or domains (patches) on length scales
$l_1..l_j..l_N$ from a smallest size $l_1=\eta$ to the system size
$l_N=L$. ``Length scale" in this more general sense means
``appropriate characteristic measure" i.e. length in one
dimension, area in two dimensions or volume in three dimensions.
In a dynamical out of equilibrium system, such as a sandpile or a
forest fire model, a steady state is achieved by driving on the
smallest length scale $l_1=\eta$ and by means of open boundaries,
removing structures on the system size $L$. In section 2 we
considered a system driven on the largest scale $L$ and
dissipating on the smallest, mapping onto fluid turbulence in a
closed system. In a model for a ferromagnet, the system may
fluctuate about an equilibrium, but nevertheless has a minimum
patch size (one spin), a maximum patch size (the system size), and
scaling of patches in between.

The global quantity $\varepsilon$ is now taken to be associated
with the total number of instantaneously active sites within each
patch. In a model realized numerically, such as a forest fire or
avalanche model, instantaneously active sites are those seen at a
given timestep in the computation. In a forest fire model, active
sites correspond to burning
 trees, in an avalanche model, to relaxing sites in evolving
avalanches \cite{bramprl}.  The global quantity may refer to the
energy dissipated by these sites, or simply refer to the time
evolution of their spatial distribution as in the case of space
occupied by anisotropic particles settling under gravity or
magnetization of spins in a ferromagnet \cite{bramprl}.

The common feature of these systems is that at any instant in time
there will be $m_j(t)$ patches on any length scale $l_j$ and
associated with each patch, $\epsilon_j^*$ of this quantity. On
each length scale $l_j$ we then have
$\epsilon_j=m_j(t)\epsilon_j^*$ and in total, $\varepsilon$ given
by (\ref{esum0}). We now assert that in common with the turbulence
model, the $\epsilon_j^*$ are independent and have intermittent,
scaling statistics (\ref{scale}). We can then envisage the
following generic scaling system which comprises:

{\em (I) Non space filling, intermittent patches:} The details of
$ <m_j^q>$ depend on the system, for example the probability of
patches $l_{j-1}$ merging, and/or patches $l_{j+1}$ breaking up to
form patches on $l_j$. We take as a necessary condition of scaling
that the moments obey:
\begin{equation}
<m_j^q>l_j^{\gamma(q)}=<m_{j-1}^q>l_{j-1}^{\gamma(q)}=<m_N^q>L^{\gamma(q)}\label{fracm}
\end{equation}
If the system were space filling, $\gamma(1)$ would be $1$  so
$\gamma(1)<1$ implies non-space filling patches. Allowing
$\gamma=\gamma(q)$ permits intermittency.

{\em (II) Fractal support:} On any patch there will be a density
of active sites $\epsilon^*_j/l_j$ which in general can vary with
$l_j$; for a system which is scaling we can however take:
\begin{equation}
\frac{\epsilon_j^*}{l_j^\alpha}=\frac{\epsilon^*_{j-1}}{l_{j-1}^\alpha}=\frac{\epsilon^*_N}{L^\alpha}\label{fracq}
\end{equation}
where $\alpha=1$ is the special case of uniform density on all
patches, and patches that do not have fractal boundaries.

{\em (III) Conservation:} Scaling implies that there is no
preferred $l_j$ on which the active sites accumulate so that the
mean will be just the ensemble average determined on any length
scale. This is consistent with conservation of active sites when
patches merge ($l_j\rightarrow l_{j+1}$) or break up
($l_j\rightarrow l_{j-1}$).

 It follows from (I), (II) and (III):
\begin{equation}
<\epsilon_j^q>
=(\epsilon_N^*)^q<m_N^q>\left(\frac{l_j}{L}\right)^{(\alpha q
-\gamma(q))}
 =\epsilon_0^q\left(\frac{l_j}{L}\right)^{-\mu(q)}\label{scale2}
\end{equation}
The condition $<\epsilon_j>=\epsilon_0$ fixes $\gamma(1)=\alpha$
or $\mu(1)=0$. The details of the system specify $\mu(2)$ which
then fixes the
 standard deviation of (\ref{kisq}) expressed through (\ref{scale2})
 and immediately leads to (\ref{sigj}).

{\em (IV) Finite size}: We finally specify the number of length
scales $N$; given scaling, a choice is constant
$(l_j/l_{j-1})=\lambda^\frac{1}{3}$ so that
$\lambda^N=(L/\eta)^3$.

 In summary then, this scaling ansatz is
that $<\epsilon_j>=\epsilon_0$, $<\epsilon_j^2>
 =\epsilon_0^2\left(l_j/L\right)^{-\mu(2)}$ with $\mu(2)\neq 0$ and
$(l_j/l_{j-1})^3=\lambda$. Any system that is specified by this
anzatz and is well approximated by (\ref{pade}) will share the
same behavior (\ref{gumb}) in the statistics of global activity
$P(\epsilon)$ that we have calculated above for the turbulence
model. Importantly, these conditions may apply to more than one
quantity in a given system, and any such quantity will share these
same statistics.

For a given system, the curve (\ref{gumb}) is specified by $a_g$
which is a function of the system parameters $N$, $\mu(2)$ and
$\lambda$. This family of curves is however insensitive to $a_g$
\cite{chapnpg}. This, combined with the practical difficulty of
obtaining good statistical resolution over fluctuations ranging
over several orders of magnitude suggests a straightforward reason
for the close, but not exact, curve collapse that has been
reported in figure 2 of \cite{bramprl}. Importantly, we do not
extend this argument to the 2D X-Y model; rather in this case  we
have utilized the correspondence of (7) with the result of [17]
(equation (27)).

 \section{Summary.}

From the starting point of a model for fluid turbulence in a
finite sized system, previously treated semi- numerically by
\cite{portellipp}, we have analytically derived the functional
form of the PDF of global energy flow in the system. This yielded
the
 dependence of its single control parameter $a_g$ on the intermittency parameter, the ratio
 between lengthscales, and the smallest and largest scale lengths
 in the system (i.e. the Reynolds number). We then directly identified this
 function with that previously obtained for fluctuations in total magnetization in the 2D
 X-Y model and thus elucidated the origin of the previously identified value $a_g \sim
 \pi/2$ \cite{bramprl}.
 The PDF was shown to
be relatively insensitive to variations
 in  $a_g$, explaining the previously reported close
 correspondence of these curves for the turbulence model and the
 2D X-Y model \cite{bramnature,portellipp}.

 Importantly, the functional form of the PDF that we derive is
 just
 that of
 the sums of a large but finite sets of
 independent numbers drawn from PDF with moments that are
 scaling. This corresponds to a model of intermittent turbulence
 in
 which one also envisages a cascade, but the cascade property is not
 intrinsic to the calculation. We suggest that this system is
 rather generic and may
encompass the wide variety
 of systems which have also been previously identified as
 exhibiting  the same functional form  for the fluctuation PDF \cite{bramprl}.

 \ack  SCC acknowledges the Radcliffe Institute for Advanced Study,
Harvard, and the PPARC for support. GR acknowledges a Leverhulme
Emeritus Fellowship. NWW thanks
 F. McRobie for assistance.

 \end{document}